\documentclass[preprint,12pt,authoryear]{elsarticle}
\newcommand{\apj}{ApJ}

\newcommand{\aap}{A\&A}
\newcommand{\apjs}{ApJS}
\newcommand{\mnras}{MNRAS}
\newcommand{\araa}{ARA\&A}

\usepackage[colorlinks=true, linkcolor=blue, citecolor=red, urlcolor=blue]{hyperref}
\usepackage{natbib}
\bibpunct{(}{)}{;}{a}{}{,}

\journal{New Astronomy Reviews}

\begin{document}

\begin{frontmatter}

\title{Comparing H$\beta$ Line Profiles in the 4D Eigenvector 1 Context}

\author{J. W. Sulentic}
\address{Instituto de Astrof\'{\i}sica de Andaluc\'{\i}a, CSIC, Apdo.
3004, 18080 Granada, Spain}
\author{P. Marziani}
\address{INAF, Osservatorio Astronomico di Padova, Vicolo dell' Osservatorio 5, 35122 Padova, Italy}
\author{S. Zamfir}
\address{Department of Physics and Astronomy, University of Alabama, Tuscaloosa, AL 35487,
USA}

\begin{abstract}

We describe a 4D Eigenvector 1 (4DE1) space that serves as a
surrogate H-R diagram for quasars. It provides a context for
describing and unifying differences between all broad line AGN.
Quasar spectra can be averaged in a  non-random way using 4DE1 just
as stellar spectra can be averaged non-randomly within the OBAFGKM
classification sequence. We find that quasars with FWHM H$\beta$
less than (Population A) and greater than (Population B) 4000 km
s$^{-1}$ show many significant differences that may point to an
actual dichotomy. Broad line profile measures and fits reenforce the
idea of a dichotomy because they are fundamentally different: Pop.A
- Lorentzian-like and Pop.B - double Gaussian. The differences have
implications both for BH mass estimation and for inferences about
source structure and kinematics.

\end{abstract}

\end{frontmatter}

\section{Introduction}
\label{intro}

Studies of quasar spectra have fallen out of fashion over the past
ten years. This impression is supported by the small number of
citations involving optical/UV spectroscopic work found from a
random NED search of citations for even the brightest (e.g., PG)
sources. One can identify at least two reasons for this paucity of
studies: 1) a widespread belief that all broad-line spectra are
essentially the same and/or 2) the impression that a deeper
understanding of broad-line phenomenology and physics is
unobtainable. The one exception, reflecting reason 1) involves
computation of mean/median composite spectra from large samples
(e.g. 2dF, SDSS). Of course if reason 1) is not correct then these
results will reenforce reason 2). Just as  the indiscriminate
averaging of OBAFGKM  stellar spectra would yield bizarre
conclusions.  The purpose of this review is to convince the reader
that reason 1) is incorrect and that understanding quasar spectral
diversity can stimulate new ideas that will remove the sense of
hopelessness reflected in reason (2).

The question is how best to represent quasar phenomenological
diversity and unite it into a coherent picture. The history of
stellar studies immediately comes to mind although quasars are
almost certainly more complicated sources. At the very least source
orientation to our line-of sight likely adds a serious complication.
We have been working on a spectroscopic unification for quasars
that, in fact, embraces all broad line emitting AGN. Gradually the
hope emerged of finding a diagnostic space capable of serving as a
surrogate H-R diagram for quasars. A 2D H-R diagram works quite well
because, among other things, stellar orientation plays no major role
in spectroscopic studies. The full power of the stellar H-R diagram
requires exploitation of more parameter dimensions. Certainly an
equivalent spectroscopic diagnostic space for quasars will require
more than two dimensions if only to remove the orientation-physics
degeneracy. In 2000 we  proposed a 4D Eigenvector 1 parameters space
(hereafter 4DE1; \citealt{Sulentic00a}). 4DE1 remains the most
promising way to emphasize the spectroscopic diversity while also
contextualizing the diverse types of broad-line emitting sources.

4DE1 space emphasizes observables with the largest intrinsic
dispersions including the most statistically significant line
profile differences. 4DE1 has roots in the Principal Component
Analysis (PCA) of the Bright (PG) Quasar Sample (\citealt{BG92};
BG92) as well as in correlations that emerged from ROSAT
\citep{Wang96}. 4DE1 in its simplest form involves two BG92
measures: 1) full width half maximum of broad H$\beta$ (FWHM
H$\beta$) and 2) equivalent width ratio of optical FeII and broad
H$\beta$: R$_{FeII}$ = W(FeII $\lambda$4750 blend)/W(H$\beta$). We
added measures of 3) the soft X-ray photon index ($\Gamma$$_{soft}$)
and 4) CIV$\lambda$1549 broad-line profile displacement at half
maximum (c(1/2)). Other points of departure from BG92 involve
comparison of radio-quiet (RQ) sources with a large radio-loud (RL)
sample. We also subordinate [OIII]$\lambda\lambda$4959,5007 measures
(although see: \citealt{Zamanov02,Marziani03a,Marziani09}). 4DE1
tells us that all quasars are not similar whether we observe them at
optical, UV or X-ray wavelengths.

\section{The Optical-UV Dimensions of 4DE1 Space}

Thus defined 4DE1 includes key measures of both representative high
(HIL: CIV$\lambda$1549, 65eV) and low (LIL: H$\beta$, 13.6eV)
ionization lines. This is important because LIL and HIL behave very
differently. While forming the core of 4DE1 the four adopted key
parameters only begin to exploit the fundamental trends and source
differences that exist. This review emphasizes results involving
4DE1 parameters describing broad emission line profiles. Figure 1
shows the latest representation of the optical plane of 4DE1
involving FWHM H$\beta$ and R$_{FeII}$ measures for the 321
brightest (g$<$17.0) low redshift (z$<$0.7) quasars (grey dots) in
the Sloan Digitized Sky Survey (SDSS; DR5) \citep{Zamfir08}. Results
are very little different if one adopts an i-band magnitude cutoff.
An i-band selection adds many nearby lower luminosity sources with
strong host galaxy contamination. SDSS does not recognize sources
with FWHM H$\beta$ $<$ 1000 km s$^{-1}$ as quasars motivating us to
add to our sample n=41 narrow-line Seyfert 1 (NLSy1) sources
brighter than g=17.5 (z$<$0.7) identified by \citet{Zhou06} (black
squares). Source occupation is very similar to what was find with
our own low redshift atlas sample of 215 sources
\citep{Marziani03b}. Figure 2 shows an optical-UV plane of 4DE1
involving FWHM H$\beta$ and CIV$\lambda$1549 profile displacement
(c(1/2)). In this case the CIV measures exhaust the HST archive of
available data with N=130 low z sources (see
\citealt{Bachev04,Sulentic07} for details).

\begin{figure}[h!]
\centering
\includegraphics[width=0.5\textwidth]{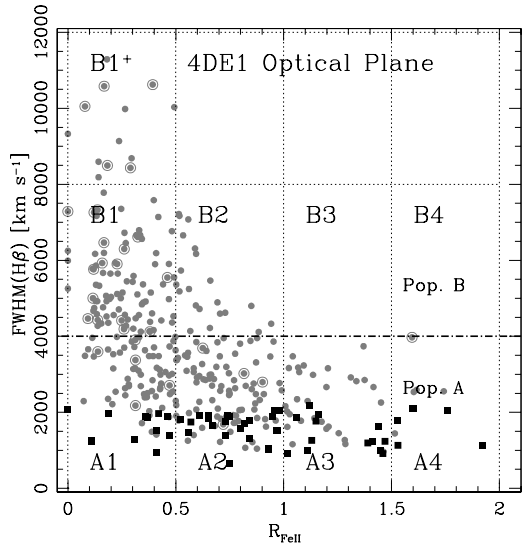}
\caption{The optical plane of the 4DE1 space. Two populations A and
B are separated at FWHM H$\beta$ = 4000 km s$^{-1}$. Bins of
$\Delta$(FWHM H$\beta$) = 4000 km s$^{-1}$ and $\Delta$(R$_{FeII}$)
= 0.5 enclose sources that have statistically similar H$\beta$ width
and FeII relative strength. Bins labels follow the convention of
\citet{Sulentic02}. Grey symbols indicate RQ sources, circled grey
symbols indicate RL quasars and the black symbols indicate the
subset of narrow broad line sources extracted from \citet{Zhou06}.}
\label{F1}
\end{figure}

\begin{figure}[h!]
\centering
\includegraphics[width=0.5\textwidth]{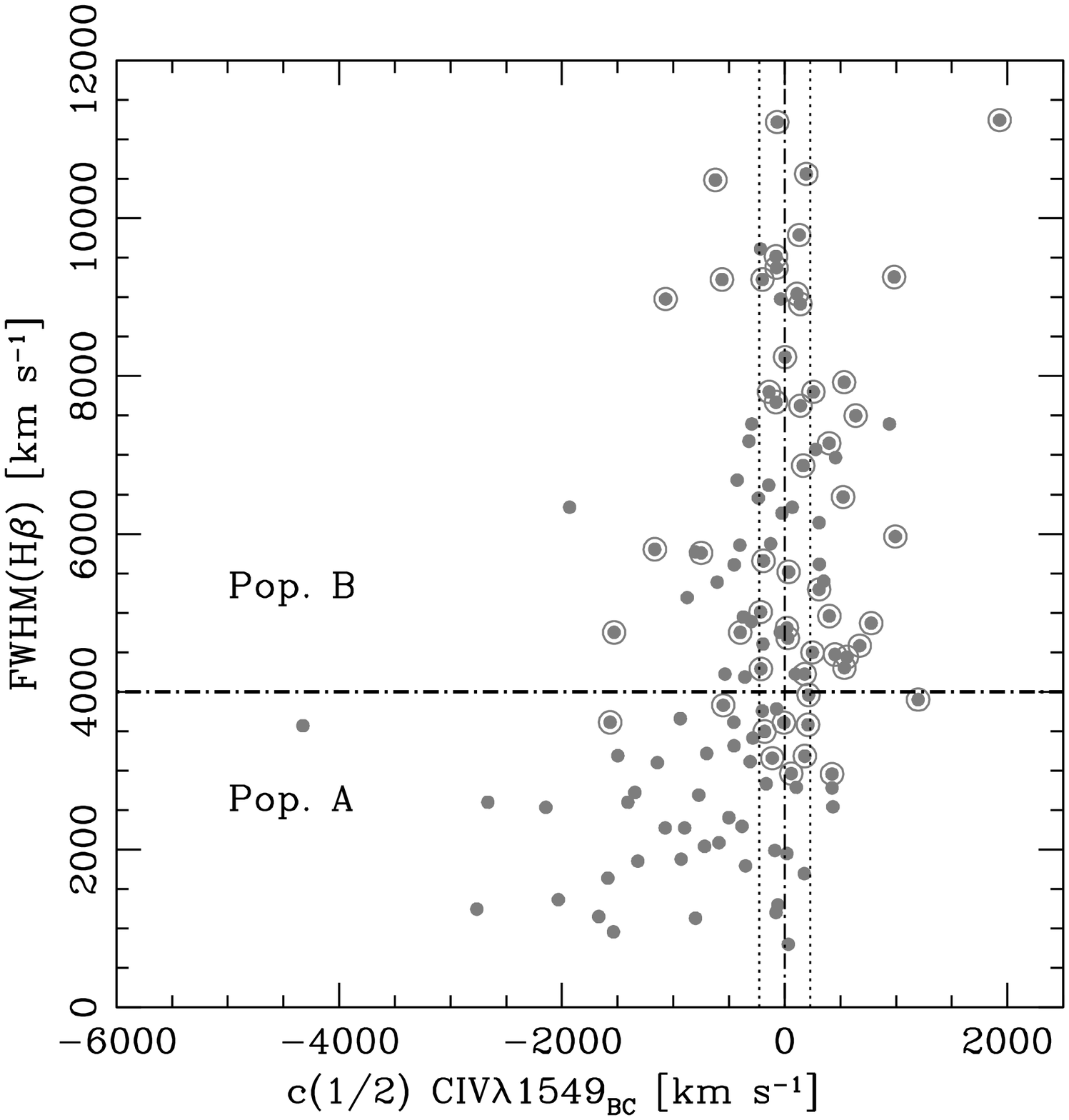}
\caption{An optical-UV plane of the 4DE1 showing the FWHM H$\beta$
versus the centroid shift of broad component of CIV$\lambda$1549 (at
1/2 fractional intensity). The symbols are similar to those in
Figure 1. The Population A/B 4000 km s$^{-1}$ boundary is also
shown.} \label{F2}
\end{figure}

Figures 1 and 2 clearly show trends but what do they mean? Does
Figure 1 show a continuous or ``main'' sequence? Or is there
evidence for an actual source dichotomy? Even if a dichotomy is not
obvious it might be useful to propose one as a means to highlight
source difference. A more striking difference is seen when RL (grey
circle-dots) and RQ (grey dots) source occupation in 4DE1 is
compared in Figure 1. This occupation difference motivates the idea
of a dichotomy because almost all RL sources show FWHM H$\beta$ $>$
4000 km s$^{-1}$ while 50-60\% of RQ sources shows smaller values.
The alternative interpretation would see RL quasars as a
fundamentally different class of quasar that partially overlaps with
RQ domain occupation in 4DE1. It really doesn't matter which
interpretation is correct, the RL-RQ difference is an important clue
about Broad Line Region (BLR) physics. Other measures (e.g., c(1/2) and $\Gamma$$_{soft}$)
point to a change in source occupation near 4000 km s$^{-1}$. Figure 2 shows that only
sources with FWHM H$\beta$ $<$ 4000 km s$^{-1}$ show a
blueshift/asymmetry of the HIL.

These empirical results further motivate the idea of contrasting
sources above and below FWHM H$\beta$ = 4000 km s$^{-1}$. The list
of source differences, and inferred physical differences, continues
to grow (see Table 5 in \citealt{Sulentic07}).

\section{Composite Spectra of Populations A and B Quasars}

Figures 3 and 4 show median composite spectra for the H$\beta$
region in the two most highly populated bins indicated in Figure 1
(A2 and B1 respectively). The composites are generated from 128 and
130 sources respectively and illustrate a striking H$\beta$ line
profile difference between Pop. A and B sources. This is an example
of averaging quasar spectra in an organized and predefined context.
It is the best way to bin the sources without imposing any
assumptions about the structure of the BLR. The size of the bins was
driven by the current accuracy of FWHM H$\beta$ and R$_{FeII}$
measures. All sources within a bin therefore have statistically
similar FWHM and R$_{FeII}$ values. We think that these median
composite spectra, or better said, the differences between them, are
the key to a deeper understanding of the quasar BLR and structural
differences between sources along the principal sequence in the 4DE1
optical plane. Figure 3 and 4 show that all quasar spectra are not
the same. Our 4000 km s$^{-1}$ boundary appears to be much more
significant than the FWHM H$\beta$ = 2000 km s$^{-1}$ boundary often
used to distinguish narrow (NlSy1) and broad (BLSy1) line Seyfert
galaxies/quasars. Comparison of sources in 1000 km s$^{-1}$
intervals above and below 2000 km s$^{-1}$ reveals little
difference; they are all what we call Pop. A sources and they are
all (similarly) different from Pop. B sources. We will consider and
contrast Figures 3 and 4 in the next two sections.

\begin{figure}[h!]
\centering
\includegraphics[width=0.5\textwidth]{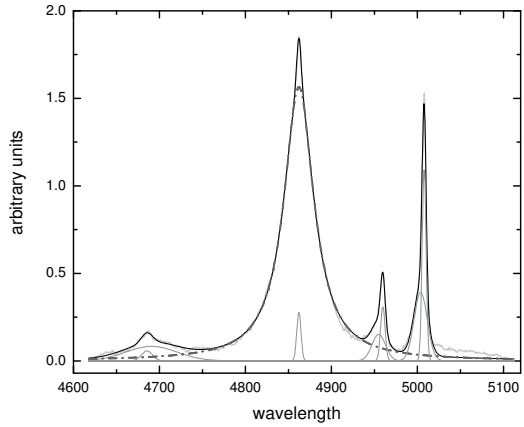}
\caption{Composite median spectrum of sources in bin A2. The
dashed-dotted profile indicates the Lorentzian fit for the broad
component of H$\beta$. The [OIII]$\lambda\lambda$ 4959,5007 and
HeII$\lambda$4686 lines require two Gaussians each.} \label{F3}
\end{figure}

\begin{figure}[h!]
\centering
\includegraphics[width=0.5\textwidth]{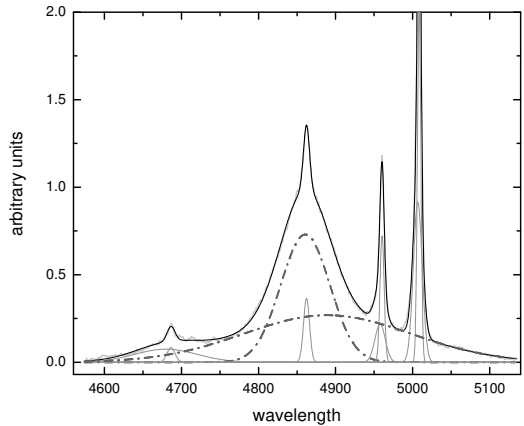}
\caption{Composite median spectrum of sources in bin B1. The
dashed-dotted profiles indicate the two Gaussians (BLR+redshifted
VBLR) for H$\beta$. The [OIII]$\lambda\lambda$ 4959,5007 and
HeII$\lambda$4686 lines are fitted with two Gaussians each.}
\label{F4}
\end{figure}

It is important to emphasize that the n$\approx$ 130 sources going
into each median composite show a wide diversity of profile shapes
although less for Pop. A than for Pop. B. Naturally we are assuming
that 4DE1 measures are first-order discriminators. We also assume
that they do not change dramatically with time. Higher-order
measures of line shift and shape are known to vary with time. One
can study these changes in an individual source via monitoring,
which costs a lot of observing time. Our alternative approach is to
generate median spectra from large enough samples that we will
likely catch source profiles in all or most significant states of
change. Median composite profiles therefore offer the possibility to
isolate the underlying most stable or common profile shape. The
shapes that we see in Figures 3 and 4 have appeared in earlier
composite spectra that we have generated with smaller and less
uniform samples. This gives us confidence that we have isolated
stable characteristics of Pop. A and B H$\beta$ line profiles.

\section{Population A Sources}

More than half ($\sim$60\%, \citealt{Sulentic00a,Zamfir08}) of a
magnitude limited quasar sample satisfies our Pop. A. definition. RL
sources are rare in Pop. A with perhaps 2-4\% satisfying standard
definitions of radio-loudness. Most RL sources show broader Balmer
line profiles and we argue that many/most formally Pop. A RL quasars
are sources that fall there because they are oriented near face-on
to our line of sight. A reasonable correction for orientation, and
assumption of a flattened BLR geometry
\citep{Marziani01,Marziani03a}, moves them into the Pop. B domain
with the bulk of RL quasars.

Figure 3 shows our Pop. A median composite for bin A2. Composites
for higher A bins would show even more extreme differences--
especially stronger R$_{FeII}$ but would have lower S/N because they
include fewer sources as is clear from Figure 1. Once we have
generated the composites we face the challenge of modeling them in
the simplest way. Over the past 10+ years, and especially for NLSy1,
it has been suggested that sources with narrower H$\beta$ profiles
can be well fit with a Lorentz function. They can then be described
as Lorentz-like profiles. Figure 3 confirms this single-component
fit for Pop. A-bin A2 sources. Whatever might be the
source-to-source differences there appears to be an underlying
stable Lorentz line. This profile shape disappears above FWHM
H$\beta$ $\approx$ 4000 km s$^{-1}$.

Figure 2 shows that CIV$\lambda$1549 profiles for Pop. A sources are
blue shifted. Blueshifts as large as 4-5000 km s$^{-1}$ are observed
although most Pop. A sources show smaller displacements of less than
500 km s$^{-1}$. While not individually significant the overall
sample displacement certainly is significant especially when
compared with the distribution of CIV line shifts in Pop. B. Pop.
A-B differences extend to all CIV measures with equivalent width
W(CIV) for Pop. A sources about 1/2 the Pop. B value. FWHM CIV shows
a smaller range than Pop. B and we find no correlation between FWHM
H$\beta$ and FWHM CIV measures for Pop. A. \citep{Sulentic07}. The
large differences between H$\beta$ and CIV profiles tell us that the
LIL and HIL in Pop. A sources arise in regions with different
geometry and kinematics.

The high S/N and resolution of this SDSS-based sample enable us to
quantify line asymmetry and shift properties of broad H$\beta$ (see
\citealt{Zamfir09} for definitions of these measures). The
distribution of profile asymmetries shows mean/median values
consistent with zero asymmetry with a sample standard deviation of
$\pm$0.11 (estimated 2$\sigma$ measurement uncertainties are
$\pm$0.16). Mean line shift measures at FW1/4 ``base'' and FW3/4
``peak'' are -42 and -41 km s$^{-1}$ respectively. A small fraction
($\sim$5\%) of Pop. A sources show a weak blue wing that involves
sources in bins A3 and A4. These tend to involve sources with the
largest CIV blueshifts \citep{Zamfir09} suggesting that we may be
detecting weak  H$\beta$ emission from the same wind or outflow
responsible for the HIL blueshifts.

The bulk of Pop.A H$\beta$ emission involves profiles that are
symmetric, unshifted and Lorentz-like. The soft X-ray excess from
these sources has been argued to be a thermal disk signatures in
what are thought to be the highest accreting sources. Similarly the
CIV blueshift/asymmetry has been ascribed to a disk wind or outflow.
Pop. A sources also lie at the high electron density n$_{e}$ end of
the 4DE1 sequence shown in Figure 1 (see also \citealt{Marziani01})
with estimated values in the range log(n$_{e}$[cm$^{-3}$]) = 10-11. All of
these results are consistent with the hypothesis that the most
important BLR component in Pop. A sources involves a thin/slim
accretion disk with n$_{e}$ and column densities high enough to
explain the strong FeII emission in these sources.

Coupled with the H$\beta$ line profile properties lead us to
conclude that FWHM H$\beta$ is a reliable virial estimator for Pop.
A quasars while FWHM CIV is not \citep{Bachev04,Sulentic07}.
Estimated black hole masses for Pop. A sources lie in the range
log(M$_{BH}$[M$_{odot}$])=6.5-8.5 with inferred Eddington ratios in the range log
L$_{BOL}$/L$_{EDD}$= 0.1-3.0. The apparent super Eddington radiators can all
be corrected to values below 0.9 by assuming that some Pop. A
radiators are observed with the line emitting accretion disk
oriented near face-on. Reasonable orientation corrections
\citep{Marziani03a} increase the smallest M$_{BH}$ estimates to
values above 7.0 thus reducing extreme L$_{BOL}$/L$_{EDD}$ values to the
sub-Eddington range.

\section{Population B Sources}

A description of population B sources will be different from the
above pop. A description in virtually every measurable parameter.
This includes emission line profiles and a comparison of Figures 3
and 4 makes it clear. Pop. B includes the bulk of the RL quasars and
about 25\% of the RQ sources. The probability of radio-loudness in
Pop. B is about 4-5 times higher than for Pop. A
\citep{Marziani03a,Zamfir08}. The meaning of this RQ-RL overlap
region in 4DE1 is unclear but there are two obvious possibilities:
1) RL sources represent a distinct quasar class that partially
overlaps the RQ domain or 2) about 25\% of RQ quasars show geometry
and kinematics identical to RL quasars. They might be RL pre- and/or
post-cursors or their central geometry may be the same as RL
galaxies, but either host galaxy morphology or black hole spin
prevents their becoming RL.

Figure 4 shows our median composite for bin B1 marked in Figure 1.
It represents a median composite of about 130 quasars. The S/N of
the Figure 3 and 4 spectra are similar, about 130. The median source
luminosity in the two bins log(L$_{BOL}$[erg s$^{-1}$])= 45.7 are
also the same (median redshifts 0.30 and 0.28 for bins A2 and B1
respectively). It is immediately obvious that the bin B1 H$\beta$
profile cannot be fit with a Lorentz function. Previous studies with
our atlas sample \citep{Marziani03b} required a minimum of two
Gaussian components to model the H$\beta$ profile after FeII
subtraction, unshifted BLR + redshifted Very Broad component VBLR
\citep{Sulentic02}. Figure 4 shows that the double Gaussian fit
again matches well the H$\beta$ profile in our SDSS composite
spectrum. If we separate Pop. B sources into luminosity bins we find
that FWHM H$\beta$ increases with source luminosity from 5200 km
s$^{-1}$ (log(L$_{BOL}$[erg s$^{-1}$])=43-44) to 9300 km s$^{-1}$
(log(L$_{BOL}$[erg s$^{-1}$])= 48-49) \citep{Marziani09}. If we
correct for the VBLR component FWHM H$\beta$$\approx$10$^4$ km
s$^{-1}$ redshifted $\approx$10$^3$ km s$^{-1}$) we find that FWHM
of BLR H$\beta$ remains almost constant and in the range 4300-4800
km s$^{-1}$. Unlike Pop. A sources, Pop. B objects show CIV profiles
that closely match those of H$\beta$. A VBLR component is very
prominent in B1$^+$\ objects and necessary to obtain a reasonable
fit of the whole CIV profile.

If we are correct in assuming that this is the classical
reverberating accretion disk component then we can again compute BH
masses for Pop. B sources. The values do not come down as much as
might be expected because the sources with the largest VBLR
correction are also the most luminous. However this tells us that
the Pop. A-B difference is not driven so much by FWHM H$\beta$ but
rather by the presence of the VBLR component that is very weak or
absent in Pop. A quasars. Figure 4 represents the strongest evidence
to date for the BLR+VBLR interpretation of the Pop. B H$\beta$ line
profile. There is a widespread idea that a double peaked accretion
disk component is hidden in the profile and indeed NGC5548 the most
reverberated source shows it. If present in all or most sources this
disk component does not show up in our high S/N median composite. It
is apparently hidden by an unshifted (or slightly blueshifted) BLR
component and the redshifted VBLR. We consider it virtually
impossible that an anomaly in the FeII emission (or error in the
FeII modeling) could produce this feature. It represents a major
challenge to our ideas about structure and kinematics of the central
region in quasars.

Paralleling our discussion of Pop. A sources we note that Pop. B
quasars show no soft X-ray excess or CIV blueshift (Figure 2). If
their presence is a signature of a stable BLR under high accretion
in Pop. A sources then the BLR is much less prominent in Pop. B
sources that host an additional VBLR emitting component. Isolating
the BLR component of H$\beta$ as a virial estimator in Pop. B
suggests that BH masses are on average larger (log(M$_{BH}$[M$_{odot}$])$\approx$7-10)
and, especially, that Eddington ratio is lower (log L$_{BOL}$/L$_{EDD}$ $\approx$ -0.7 to -1.4) than in Pop. A.

\section{Meaning of a Pop. A-B Dichotomy and VBLR?}

These questions are asked together because we think they are
related. After all the appearance of the VBLR component near FWHM
H$\beta$=4000 km s$^{-1}$ is one of the strongest arguments for a
Pop. A-B dichotomy \citep{Sulentic00b,Sulentic07}. We therefore
conclude that the dichotomy is real and likely related to a critical
Eddington ratio below which BLR structure undergoes a significant
change in geometry and kinematics (see also
\citealt{Collin06,Bonning07,Kelly08,Hu08}). If a critical
L$_{BOL}$/L$_{EDD}$ is the driver then how does the VBLR relate to this
change? So far we can infer four VBLR characteristics: 1) it is very
broad, 2) it is highly redshifted, 3) it is reasonably well fit by a
symmetric Gaussian, and 4) it becomes more prominent at higher
luminosity (due to a weakening of the central BLR component). The
equivalent width of the VBLR component is roughly constant over six
decades of luminosity, which implies production in optically-thick,
photoionized gas. We can identify three classes of models that might
explain the VBLR: a) gravitation (clues 1 and 2), b) infall (clue
2), c) photon downshifting through scattering processes (clues 1, 2
and 4). Inferred BH mass values are not large enough to make it
likely that the VBLR arises close enough to the BH and clue 3 argues
against this interpretation as well. We think that b) or c) likely
provides the solution and we encourage to pursue them just as we are
doing.



\begin{thebibliography}{00}
\footnotesize
\bibitem[Bachev et al.(2004)]{Bachev04} Bachev, R., Marziani, P., Sulentic, J. W., Zamanov, R., Calvani, M., Dultzin-Hacyan, D. 2004 \apj, 617, 171
\vspace*{-1.1em}
\bibitem[Bonning et al.(2007)]{Bonning07} Bonning, E. W., Cheng, L., Shields, G. A., Salviander, S., Gebhardt, K. 2007, \apj, 659, 211
\vspace*{-1.1em}
\bibitem[Boroson \& Green(1992)]{BG92} Boroson, T. A. \& Green, R. F. 1992, \apjs, 80, 109
\vspace*{-1.1em}
\bibitem[Collin et al.(2006)]{Collin06} Collin, S., Kawaguchi, T., Peterson, B. M., Vestergaard, M. 2006, \aap, 456, 75
\vspace*{-1.1em}
\bibitem[Hu et al.(2008)]{Hu08} Hu, C., Wang, J.-M., Ho, L. C., Chen, Y.-M., Zhang, H.-T., Bian, W.-H., Xue, S.-J. 2008, \apj, 687, 78
\vspace*{-1.1em}
\bibitem[Kelly et al.(2008)]{Kelly08} Kelly, B. C., Bechtold, J., Trump, J. R., Vestergaard, M., Siemiginowska, A. 2008, \apjs, 176, 355
\vspace*{-1.1em}
\bibitem[Marziani et al.(2001)]{Marziani01} Marziani, P., Sulentic, J. W., Zwitter, T., Dultzin-Hacyan, D., Calvani, M. 2001, \apj, 558, 553
\vspace*{-1.1em}
\bibitem[Marziani et al.(2003a)]{Marziani03a} Marziani, P., Zamanov, R. K., Sulentic, J. W., Calvani, M. 2003a, \mnras, 345, 1133
\vspace*{-1.1em}
\bibitem[Marziani et al.(2003b)]{Marziani03b} Marziani, P., Sulentic, J. W., Zamanov, R., Calvani, M., Dultzin-Hacyan, D., Bachev, R., Zwitter, T. 2003b, \apjs, 145, 199
\vspace*{-1.1em}
\bibitem[Marziani et al.(2009)]{Marziani09} Marziani, P., Sulentic, J. W., Stirpe, G. M., Zamfir, S., Calvani, M. 2009, \aap, 495, 83
\vspace*{-1.1em}
\bibitem[Sulentic et al.(2000a)]{Sulentic00a} Sulentic, J. W., Zwitter, T., Marziani, P., Dultzin-Hacyan, D. 2000a, \apj, 536, L5
\vspace*{-1.1em}
\bibitem[Sulentic et al.(2000b)]{Sulentic00b} Sulentic, J. W., Marziani, P., Dultzin-Hacyan, D. 2000b, \araa, 38, 521
\vspace*{-1.1em}
\bibitem[Sulentic et al.(2002)]{Sulentic02} Sulentic, J. W., Marziani, P., Zamanov, R., Bachev, R., Calvani, M., Dultzin-Hacyan, D. 2002, \apj, 566, L71
\vspace*{-1.1em}
\bibitem[Sulentic et al.(2007)]{Sulentic07} Sulentic, J. W., Bachev, R., Marziani, P., Negrete, C. A., Dultzin, D. 2007, \apj, 666, 757
\vspace*{-1.1em}
\bibitem[Wang et al.(1996)]{Wang96} Wang, T., Brinkmann, W., Bergeron, J. 1996, \aap, 309, 81
\vspace*{-1.1em}
\bibitem[Zamanov et al.(2002)]{Zamanov02} Zamanov, R., Marziani, P., Sulentic, J. W., Calvani, M., Dultzin-Hacyan, D., Bachev, R. 2002, \apj, 576, L9
\vspace*{-1.1em}
\bibitem[Zamfir et al.(2008)]{Zamfir08} Zamfir, S., Sulentic, J. W., Marziani, P. 2008, \mnras, 387, 856
\vspace*{-1.1em}
\bibitem[Zamfir et al.(2009)]{Zamfir09} Zamfir, S., Sulentic, J. W., Marziani, P.
2009, submitted to \mnras
\vspace*{-1.1em}
\bibitem[Zhou et al.(2006)]{Zhou06} Zhou, H., Wang, T., Yuan, W., Lu, H., Dong, X., Wang, J., Lu, Y. 2006, \apjs, 166, 128
\vspace*{-1.1em}

\end{thebibliography}
\end{document}